\documentclass[twocolumn,showpacs,preprintnumbers,amsmath,amssymb]{revtex4}
\usepackage{graphicx}
\usepackage{dcolumn}
\usepackage{bm}

\newcommand\beq{\begin{equation}}
\newcommand\eeq{\end{equation}}
\newcommand\bseq{\begin{subequations}}
\newcommand\eseq{\end{subequations}}
\newcommand\bfig{\begin{figure}}
\newcommand\efig{\end{figure}}

\begin{document}

\title{Classical analogue of displaced Fock states and quantum correlations in Glauber-Fock photonic lattices}

\author{Robert Keil$^{1,\ast}$, Armando Perez-Leija$^{2,3}$, Felix Dreisow$^1$, Matthias Heinrich$^1$, Hector Moya-Cessa$^3$, Stefan Nolte$^1$, Demetrios N. Christodoulides$^2$, and Alexander Szameit$^{1}$}

\affiliation{$^1$\,Institute of Applied Physics, Friedrich-Schiller-Universit\"{a}t Jena, Max-Wien-Platz 1, 07743 Jena, Germany}
\affiliation{$^2$\,CREOL/College of Optics, University of Central Florida, Orlando, Florida, USA}
\affiliation{$^3$\,INAOE, Coordinacion de Optica, A.P. 51 y 216, 72000 Puebla, Mexico}
\affiliation{$^*$robert.keil@uni-jena.de}
\date{\today}

\begin{abstract}
Coherent states and their generalisations, displaced Fock states, are of fundamental importance to quantum optics. Here we present a direct observation of a classical analogue for the emergence of these states from the eigenstates of the harmonic oscillator. To this end, the light propagation in a Glauber-Fock waveguide lattice serves as equivalent for the displacement of Fock states in phase space. Theoretical calculations and analogue classical experiments show that the square-root distribution of the coupling parameter in such lattices supports a new family of intriguing quantum correlations not encountered in uniform arrays. Due to the broken shift-invariance of the lattice, these correlations strongly depend on the transverse position. Consequently, quantum random walks with this extra degree of freedom may be realised in Glauber-Fock lattices.
\end{abstract}


\maketitle
Since their introduction by Glauber in 1963, coherent states have been the subject of extensive research within the framework of quantum optics \cite{Glauber:CoherentStates}. 
The average positions and momenta of these minimum-uncertainty wavepackets are known to follow the motion of a classical oscillator, thereby establishing an important bridge between quantum and classical mechanics \cite{Schrödinger:Wellenpakete}.
Coherent states arise either as eigenkets of the annihilation operator or from a displacement of the ground state of the quantised harmonic oscillator in phase space \cite{Glauber:CoherentStates}. In general, if displacements of the oscillator eigenstates (termed Fock states or number states) are considered, a more general class of states, so-called displaced Fock states (DFS), can be obtained \cite{Boiteux:DisplacedFockState,deOliveira:DisplacedFockStates}.
These states are of great relevance to many areas of quantum optics, the probably most important one being the direct measurement of Wigner functions \cite{Lutterbach_Banaszek}, which has been successfully performed on propagating coherent states \cite{Banaszek:WignerfunctionMeasurement}, on single photons in cavities \cite{Bertet:WignerMeasurementOnePhotonCavity} and on motional states of trapped atoms \cite{Leibfried:WignerMeasurementTrappedAtoms}. Furthermore, DFS constitute the eigenstates of Jaynes-Cummings systems with coherently driven atoms \cite{Alsing:DynamicStarkEffectJCS}, and recently, entangled DFS have been proposed for quantum dense coding \cite{Podoshvedov:DisplacedFockStateDenseCoding}. DFS have been successfully generated by superposing a Fock state with a coherent state on a beam splitter \cite{Lvovsky:DisplacedFockStateGeneration}. However, due to the difficulties in generating pure Fock states of higher orders, this approach is limited to the lowest-order DFS. To our knowledge, a direct observation of the genesis of these states has also not been possible to date.\\
Quite recently, an optical system has been proposed which allows for a direct observation of a classical analogue for the displacement of Fock states \cite{Perez-Leija:GlauberFockLattice}: A photonic lattice of evanescently coupled waveguides \cite{Christodoulides:DiscretizingLight}, with a square-root distribution of the coupling between adjacent guides. In these Glauber-Fock photonic lattices, every excited waveguide represents a Fock state and the spatial evolution of the light field corresponds to the probability amplitudes of the DFS in the number basis. Thereby, the emergence of these fundamental states and the underlying displacement process can be directly visualised. As no collapse of the wavefunction occurs for classical light, the displacement can be observed for a wide range of displacement amplitudes simultaneously.\\
In this letter, we present the first experimental realisation of a Glauber-Fock photonic lattice and directly observe the classical analogue of Fock state displacements up to the fourth oscillator eigenstate. This arrangement is implemented by direct femtosecond (fs) laser waveguide inscription in fused silica \cite{Szameit:Coupling}, whereby the required coupling distribution is achieved by a controlled variation of the distance between neighbouring waveguides.\\
From another perspective, Glauber-Fock lattices provide a fertile ground for quantum random walks (QRWs) of correlated particles, as we show in the second part of this letter. If pairs of identical non-interacting photons are launched into a uniform waveguide lattice, the quantum interference of all possible paths leads to correlation of the photons \cite{Bromberg:PhotonCorrelationArray1D}, thereby enabling continuous-time correlated QRWs in a state space much larger than for a single photon \cite{Peruzzo:QuantumWalkPhotonPairs}. The correlation of identical particles has also been analysed for disordered lattices exhibiting Anderson localisation \cite{Lahini:CorrelationsAnderson} and for Bloch oscillations occurring in lattices with a linear gradient in their propagation constant \cite{Bromberg:PhotonCorrelationBloch}.
As all these lattices are shift-invariant and infinite in the transverse dimension, the possible trajectories in a QRW are independent of their starting point. Breaking this invariance and introducing a boundary can thus embed the additional degree of freedom of transverse position into the QRW. We therefore analyse the correlations for pairs of separable and path-entangled photons as well as of fermions in a semi-infinite Glauber-Fock lattice and show how their correlation patterns uniquely depend on the input position.

In order to observe the displacement of Fock states in the optical domain, one requires a lattice of  single-mode waveguides, whose coupling coefficients between adjacent elements vary with the square-root of the site labelling index $\rm{n}$ \cite{Perez-Leija:GlauberFockLattice}:
\beq
i\frac{d\phi_{\rm{n}}}{dz}+C_{\rm{n}}\phi_{\rm{n-1}}+C_{\rm{n+1}}\phi_{\rm{n+1}}=0;\;C_{\rm{n}}=\sqrt{\rm{n}}C_1.
\label{coupling}
\eeq
Thereby, $\phi_{\rm{n}}$ denotes the modal field amplitude in guide $\rm{n}$, $z$ is the longitudinal coordinate and $C_1$ is the coupling strength between the first two guides (see Fig. \ref{Fig1}). We represent the oscillator eigenstates by single waveguides, i.e., the Fock state $\left|\rm{k}\right\rangle$ shall correspond to the situation when only guide $\rm{k}$ is excited: $\phi_{\rm{n}}=\delta_{\rm{nk}}$.
If light is launched into this single site at $z=0$, the light propagation along $z$ will map the displacement of the Fock state $\left|\rm{k}\right\rangle$ along the imaginary axis of the quadrature phase space. More specifically, the field amplitudes evolve analogous to the matrix elements of the unitary displacement operator $\hat{D}(\alpha)=\exp(\alpha a^{\dagger}-\alpha^{\ast}a)$ \cite{Glauber:CoherentStates,Perez-Leija:GlauberFockLattice}:
\beq
\phi_{\rm{n}}(z)=\left\langle \rm{n}\right|\hat{D}(iC_1z)\left|\rm{k}\right\rangle,
\label{displacement}
\eeq
with $a^{(\dagger)}$ being the ladder operators of the oscillator: $a\left|\rm{n}\right\rangle=\sqrt{\rm{n}}\left|\rm{n-1}\right\rangle$, $a^{\dagger}\left|\rm{n}\right\rangle=\sqrt{\rm{n+1}}\left|\rm{n+1}\right\rangle$. Hence, after propagating a distance $z$, the light intensity distribution $|\phi_{\rm{n}}|^2=\left|\left\langle\rm{n}\right|\hat{D}(iC_1z)\left|\rm{k}\right\rangle\right|^2$ is equivalent to the number distribution of the DFS $\hat{D}(iC_1z)\left|\rm{k}\right\rangle$ \cite{deOliveira:DisplacedFockStates}.
\bfig
\centering
\includegraphics[width=80mm]{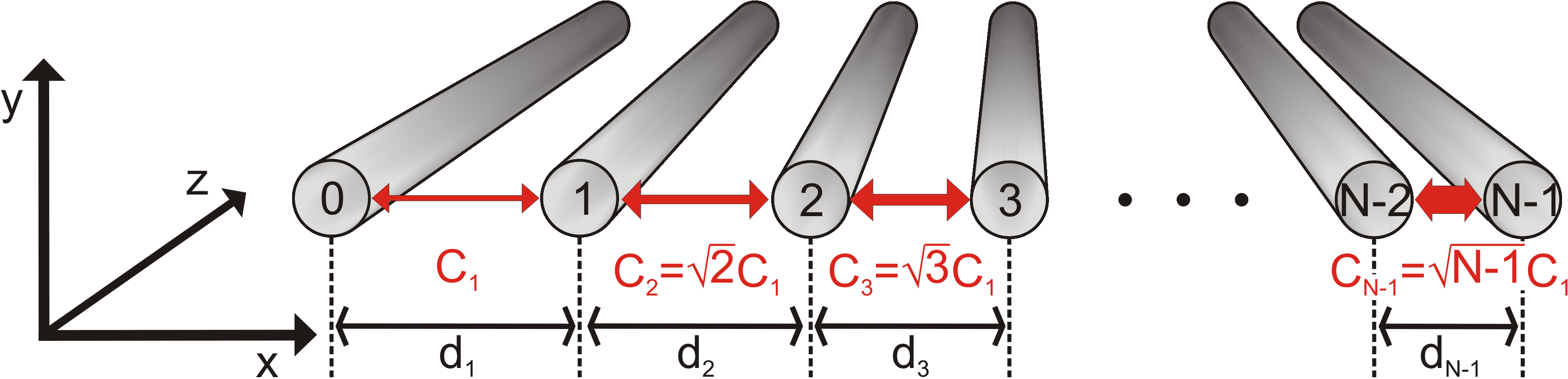}
\caption{\label{Fig1}(Color Online) Schematic view of a Glauber-Fock lattice of $N$ waveguides.}
\efig
\begin{figure*}
\centering
\includegraphics[width=160mm]{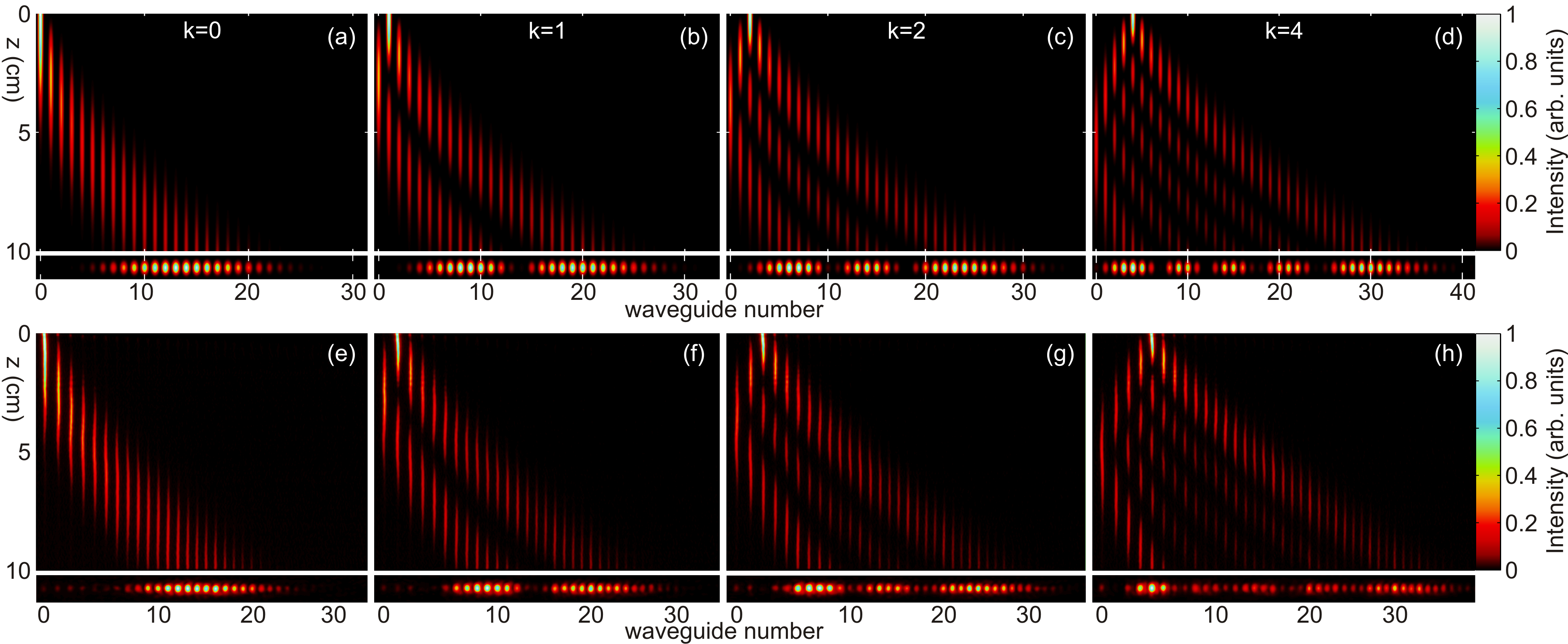}
\caption{\label{Fig2}(Color Online) Light propagation in a Glauber-Fock lattice. (a-d) Calculated intensity evolution and output intensity profiles for the input sites (a) $\rm{k}=0$, (b) $\rm{k}=1$, (c) $\rm{k}=2$ and (d) $\rm{k}=4$ representing the displacement of the Fock states $\left|0\right\rangle$, $\left|1\right\rangle$, $\left|2\right\rangle$ and $\left|4\right\rangle$, respectively. (e-h) Experimental fluorescent images of the intensity evolution and nearfield images of the output facet for a single-waveguide excitation of these sites with $\lambda=633\rm{nm}$. All images have been normalised to their respective peak intensity. For generality, the theoretical images are shown for equidistant waveguides, while in the experiments the sites are distributed according to $d_{\rm{n}}=d_1-\kappa\log(\sqrt{\rm{n}})$.}
\end{figure*}

In the weak coupling regime, $C_n$ depends exponentially on the distance between the guides: $C_{\rm{n}}=C_1\exp[-(d_{\rm{n}}-d_1)/\kappa]$, with $\kappa$ and $d_1$ being fit parameters, if $C_1$ is predetermined \cite{Szameit:Coupling}. Consequently, the coupling dependence of Eq.~(\ref{coupling}) is readily achieved by inscribing the waveguides with $d_{\rm{n}}=d_1-\kappa\log(\sqrt{\rm{n}})$ as the distribution of separations.
Measuring the dependence of coupling on waveguide separation at a wavelength of $\lambda=633\rm{nm}$, we found the parameters $d_1=23\rm{\mu m}$ and $\kappa=5.5\rm{\mu m}$ for a desired coupling of $C_1=0.37\rm{cm^{-1}}$. Using these results, we inscribed a Glauber-Fock lattice with $N=59$ waveguides and $10\rm{cm}$ length in fused silica, corresponding to a maximum displacement amplitude $\alpha=3.7i$. We employed fluorescence microscopy to directly observe the intensity evolution of the injected light \cite{Szameit:Coherence} and imaged the output intensity patterns onto a CCD camera. Figs. \ref{Fig2}(a-d) present numerical results obtained from integrating Eq.~(\ref{coupling}) for several different input sites while the experimental data is shown in Figs. \ref{Fig2}(e-h), matching the theoretical expectations very closely. The results clearly map the genesis of coherent states with their typical Poisson distribution from the displacement of the ground state of the oscillator (a, e) as well as the generation of higher-order DFS from its higher eigenstates (b-d, f-h). The intensity distribution features $\rm{k+1}$ maxima when site $\rm{k}$ is excited, in agreement to the characteristic oscillation with $\rm{k+1}$ maxima in the number distribution of the DFS $\hat{D}(\alpha)\left|k\right\rangle$. Interestingly, these oscillations are analogous to the oscillations of the Franck-Condon factors, governing the transition probabilities between vibrational levels during an electronic transition in molecules \cite{deOliveira:DisplacedFockStates,FranckCondon:VibronicTransitions}. In that context, the DFS correspond to the vibrational levels of the excited state, while the levels of the electronic ground state are undisplaced oscillator eigenstates.
Note, that the right-hand boundary at $\rm{n}=N-1$ is not reached by the propagating light. 
Thus, the lattice can be considered as effectively semi-infinite, corresponding to the semi-infinite set of Fock states. The largest coupling reported for laser written waveguide arrays is $C_{N-1}\propto5.5\rm{cm}^{-1}$ \cite{Dreisow:BlochZenerSuperlattice}, which limits the maximum number of waveguides for the given $C_1$ to $N\propto200$.\\
As these measurements demonstrate, Glauber-Fock lattices support an optical emulation of DFS for a wide range of parameters, providing direct insight into their generation characteristics. The success of such an optical emulation clearly highlights the wave nature of DFS. As governed by Eq.~(\ref{displacement}), the phases of the Fock coefficients are also encoded in the modal amplitudes. Hence, a full reconstruction of the DFS could be achieved by interferometric phase retrieval at the end of the lattice, e.g., by superposing the output field with a reference wave and phase-stepping \cite{Fleischer:1DSolitonsPhotorefractive}.
\begin{figure*}
\centering
\includegraphics[width=160mm]{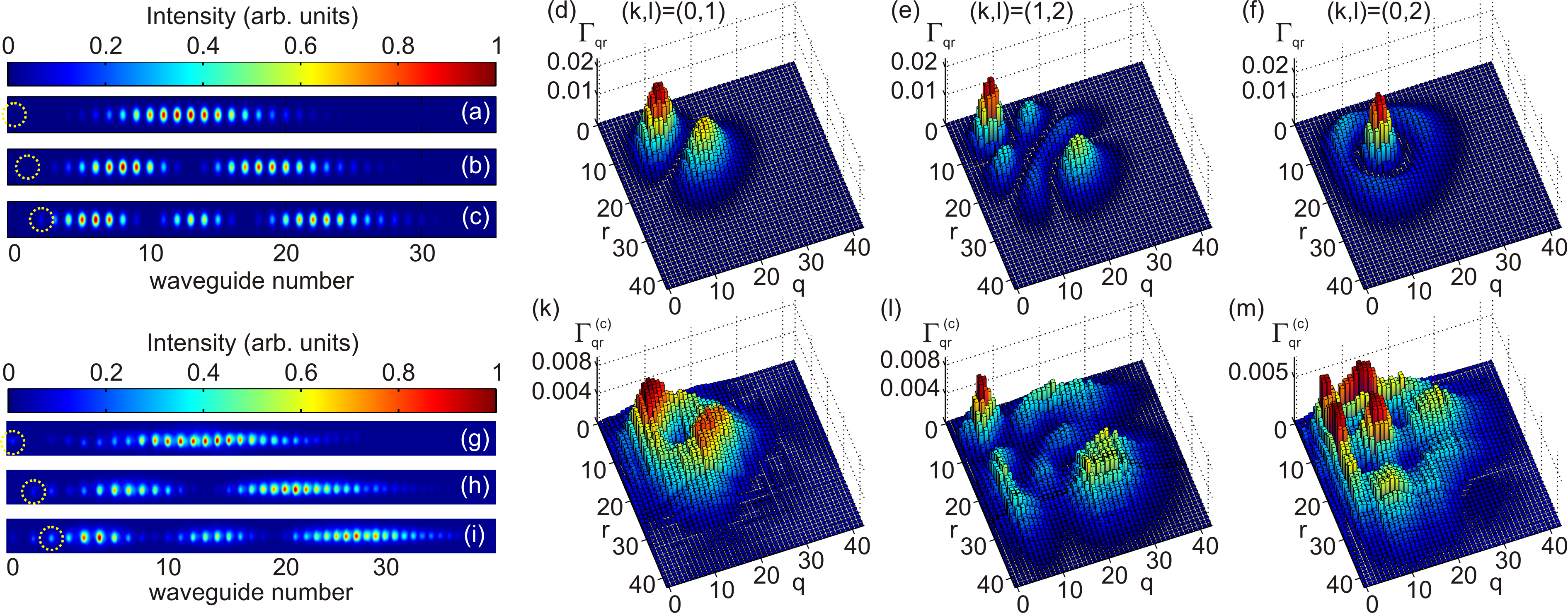}
\caption{\label{Fig3}(Color Online) Photon correlations in a Glauber-Fock lattice. (a-c) Calculated output intensities for the input sites (a) $\rm{k}=0$, (b) $\rm{k}=1$ and (c) $\rm{k}=2$. (d-f) Calculated photon correlation for the input $\left|\Psi_0\right\rangle=a_{\rm{k}}^{\dagger}a_{\rm{l}}^{\dagger}\left|\emptyset\right\rangle$ with (d) $\rm{\left(k,l\right)}=\left(0,1\right)$, (e) $\left(\rm{k,l}\right)=\left(1,2\right)$ and (f) $\left(\rm{k},\rm{l}\right)=\left(0,2\right)$. (g-i) Measured classical output intensities at $\lambda=800\rm{nm}$. (k-m) Measured classical estimates. All images have been normalised to their respective peak values.}
\end{figure*}
\bfig
\centering
\includegraphics[width=80mm]{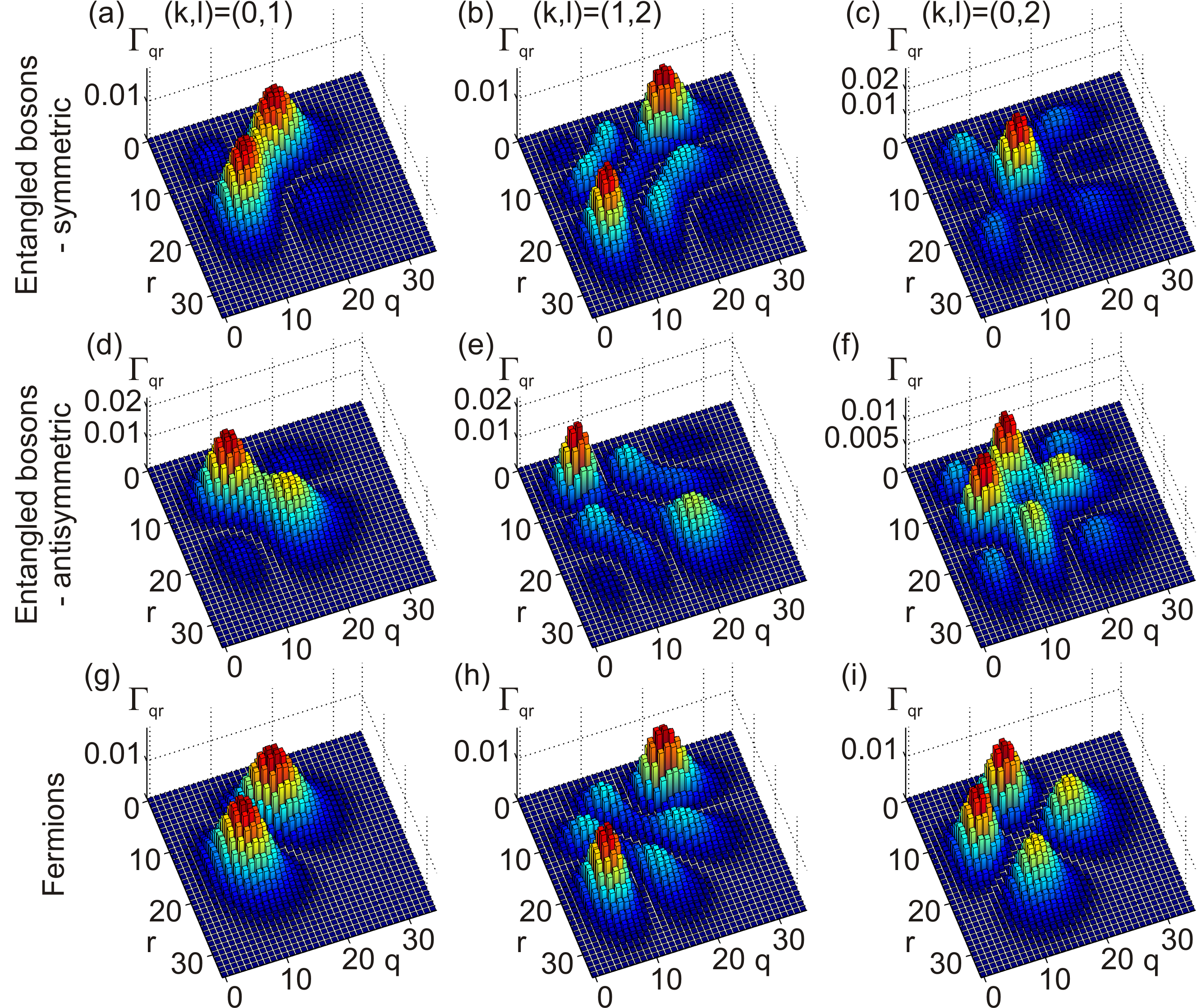}
\caption{\label{Fig4}(Color Online) Expected correlation patterns of path-entangled photons and separable fermions. (a-c) Photon correlation for the symmetric boson input state $\left|\Psi_0^+\right\rangle$. (d-f) Same for the antisymmetric boson state $\left|\Psi_0^-\right\rangle$. (g-i) Correlation function of two fermions being launched into sites $\rm{k}$ and $\rm{l}$.}
\efig 

Let us now turn to QRWs of correlated photons in such Glauber-Fock lattices. As evident from Fig.~\ref{Fig2}, the intensity distribution of a classical light field is highly characteristic for its input waveguide. Hence, the same feature applies to the output probability distribution of a single photon. Figs.~\ref{Fig3}(a-c) show the expected distribution for a single photon excitation of the first three sites. If two indistinguishable photons propagate in the lattice, all possible paths will interfere, and one can therefore expect correlation patterns which are unique for each combination of input positions, in contrast to the correlations arising in uniform lattices where all waveguides are embedded in identical coupling environments \cite{Bromberg:PhotonCorrelationArray1D,Peruzzo:QuantumWalkPhotonPairs}.\\
At first, we consider input states of separable photons, where two photons are launched into the waveguides $\rm{k}$ and $\rm{l}$: $\left|\Psi_0\right\rangle=a_{\rm{k}}^{\dagger}a_{\rm{l}}^{\dagger}\left|\emptyset\right\rangle$. Here, $a_{\rm{k}}^{(\dagger)}$ denotes the boson annihilation (creation) operator for a photon in guide $\rm{k}$ and $\left|\emptyset\right\rangle$ is the vacuum state. The probability of coincident detection of photons in guides $\rm{q}$ and $\rm{r}$ is determined by the photon number correlation \cite{Bromberg:PhotonCorrelationArray1D}:
\beq
\Gamma_{\rm{qr}}=\left\langle a_{\rm{q}}^{\dagger}a_{\rm{r}}^{\dagger}a_{\rm{r}}a_{\rm{q}}\right\rangle=\left|U_{\rm{qk}}U_{\rm{rl}}+U_{\rm{ql}}U_{\rm{rk}}\right|^2,
\label{bosoncorr}
\eeq
with $U_{\rm{mn}}=\left(\rm{e}^{iz\cal{C}}\right)_{\rm{mn}}$ and the coupling matrix being  ${\cal{C}}_{\rm{mn}}=C_1\left(\sqrt{\rm{n+1}}\delta_{\rm{mn+1}}+\sqrt{\rm{n}}\delta_{\rm{mn-1}}\right)$ for the Glauber-Fock lattice.
We calculated $\Gamma$ for a $10\rm{cm}$ long lattice with $C_1=0.36\rm{cm^{-1}}$ and a set of input configurations involving the first three sites [Figs.~\ref{Fig3}(d-f)]. The correlation patterns, and thereby the trajectories of a correlated QRW, show typical bosonic bunching behaviour: on-diagonal peaks, corresponding to a high probability of detecting both photons in the same output region. Unlike in uniform arrays \cite{Bromberg:PhotonCorrelationArray1D}, the correlations are unique for each input state and depend critically on its transverse position, not only on the interleaving distance among the excitation sites. This is due to the extra phase the photons acquire during reflection at the boundary as well as due to the specific Glauber-Fock coupling distribution.\\
The correlation function in any photonic lattice can be estimated with classical light by intensity correlation measurements \cite{Keil:PhotonCorrelationArray2D}. We performed such measurements with coherent light beams in a second lattice of $59$ waveguides, designed for $\lambda=800\rm{nm}$. In this spectral range the fabrication parameters were found to be $\kappa=10.7\rm{\mu m}$ and $d_1=34\rm{\mu m}$ for $C_1=0.36\rm{cm^{-1}}$. Figs.~\ref{Fig3}(g-i) show the measured output intensities from single-waveguide excitation, clearly exhibiting the typical number distribution of DFS. Evidently, they closely match the theoretical data presented in Figs.~\ref{Fig3}(a-c), apart from a slightly increased width of the light distribution, which can be attributed to the growth of the coupling being slightly stronger than the desired $\sqrt{\rm{n}}$-dependence. The classical estimates $\Gamma^{(c)}$ are obtained by launching two coherent laser beams of equal amplitude and adjustable relative phase into the sites $\rm{k}$ and $\rm{l}$, measuring the intensity correlation among the waveguides at the output, averaging over $60$ random relative phases and deducting intensity products from single-waveguide excitation. For a detailled description of the method, we refer to \cite{Keil:PhotonCorrelationArray2D} (see Eqs. (7) and (8) therein). The results obtained by this scheme are presented in Figs.~\ref{Fig3}(k-m). Despite a significant level of noise, most likely due to fluctuations of phase and amplitude of the classical light fields, the essential features of $\Gamma$ are well recovered by the classical estimate, namely the positions of its maxima and the \lq\lq valleys\rq\rq\space of destructive quantum interference. The shift towards larger waveguide indices is due to the increased spread of the light distribution as observed for single-waveguide excitation [Figs.~\ref{Fig3}(g-i)].\\
Finally, we theoretically study input states of path-entangled photon pairs, where both photons are sent into either of the guides $\rm{k}$ and $\rm{l}$ ($\rm{k}\neq\rm{l}$): $\left|\Psi_0^{\pm}\right\rangle=\frac{1}{2}(a_{\rm{k}}^{\dagger2}\pm a_{\rm{l}}^{\dagger2})\left|\emptyset\right\rangle$. These biphoton states are the lowest-order $N00N$-states, a class of maximally entangled $N$-photon states \cite{Kok:N00N-StateProposal}. Upon propagation of such a state in a photonic lattice the photon correlation yields \cite{Keil:ClassicalSchemeEntangled}: $\Gamma_{\rm{qr}}^{\pm}=\left|U_{\rm{qk}}U_{\rm{rk}}\pm U_{\rm{ql}}U_{\rm{rl}}\right|^2$.
The correlations have been calculated for the lowest order Fock states for the symmetric ($+$) as well as for the antisymmetric state ($-$) [Figs.~\ref{Fig4}(a-f)]. As before, the correlation maps are highly distinct for each particular input configuration and become more involved when higher excitation sites are used.\\ 
To gain a more thorough insight into the nature of these correlations, it is worthwile to compare them to the correlation of product states of identical bosons [Eq.~(\ref{bosoncorr})] and fermions in Glauber-Fock lattices. As recently demonstrated, fermionic correlations can be accurately simulated in birefringent waveguide lattices, provided that the two polarization states are entangled across different sites \cite{Matthews:SimulatingAnyons}. Introducing the annihilation(creation) operators $b_{\rm{n}}^{(\dagger)}$ for fermions in a lattice \cite{Negele:QuantumManyParticle} and considering initial states of the type $\left|\Psi_0^{(f)}\right\rangle=b_{\rm{k}}^{\dagger}b_{\rm{l}}^{\dagger}\left|\emptyset\right\rangle$ ($\rm{k}\neq\rm{l}$), the fermionic correlation function yields \cite{Lahini:CorrelationsAnderson}: $\Gamma_{\rm{qr}}^{(f)}=\left\langle b_{\rm{q}}^{\dagger}b_{\rm{r}}^{\dagger}b_{\rm{r}}b_{\rm{q}}\right\rangle=\left|U_{\rm{qk}}U_{\rm{rl}}-U_{\rm{ql}}U_{\rm{rk}}\right|^2$.
The results displayed in Figs.~\ref{Fig4}(g-i) show a strong anticorrelation: Off-diagonal peaks dominate the correlation map and there is zero probability to find both particles in the same channel, as required by Pauli's exclusion principle. Quite remarkably, the correlation patterns for the entangled photons [Figs.~\ref{Fig4}(a-f)] reveal a composition of bosonic bunching [compare Figs.~\ref{Fig3}(d-f)] and fermionic antibunching features, similar to the situation in disordered lattices \cite{Lahini:CorrelationsAnderson}. In some cases the main peaks follow a bosonic behaviour(c, d, e), while in others the fermionic anticorrelation prevails (a, b, f). However, as the entangled photons are still bosons, they never exhibit the strong on-diagonal zero  trench characteristic for fermions. Note that a classical estimate for the correlation of path-entangled photon pairs can be obtained from intensity correlations of two coherent light beams as well, but requires a precisely controlled relative phase \cite{Keil:ClassicalSchemeEntangled}.

In conclusion, we directly observed a classical analogue for the displacement of Fock states by monitoring the propagation of classical light waves in Glauber-Fock photonic lattices. This demonstrates that essentially wave mechanics governs the displacement process. As every waveguide is subjected to a different coupling environment, correlations of identical particles evolving in such a lattice are highly characteristic for each input configuration. Therefore, quantum random walks exploiting the additional layer of complexity associated with this degree of freedom seem in reach. 

The authors acknowledge support by Deutsche Forschungsgemeinschaft
(Research Unit 532 and Leibniz program). R.K. is supported by Abbe School of Photonics, A.S. is supported by the Leopoldina - German Academy of Science (Grant LPDS 2009-13).


\end{document}